\documentclass[
12pt]{article}
\usepackage{color}
\usepackage{graphicx}
 \usepackage{amsmath,amssymb,array,calc,rotating,epsfig,psfrag}  
\usepackage[nosort]{cite} 

\definecolor{darkgreen}{rgb}{0,0.55,0}

\newcommand{\bea}{\begin{eqnarray}}
\newcommand{\eea}{\end{eqnarray}}
\newcommand{\be}{\begin{equation}}
\newcommand{\ee}{\end{equation}}

\newcommand{\sbra}[1]{\left[ #1 \right|}

\newcommand{\ket}[1]{\left|#1\right\rangle }
\newcommand{\sket}[1]{\left| #1 \right]}

\def\revise#1       {\raisebox{-0em}{\rule{3pt}{1em}}%  
                     \marginpar{\raisebox{.5em}{\vrule width3pt\  
                     \vrule width0pt height 0pt depth0.5em  
                     \hbox to 0cm{\hspace{0cm}{%  
                     \parbox[t]{4em}{\raggedright\footnotesize{#1}}}\hss}}}}

\def\sqr#1#2{{\vcenter{\vbox{\hrule height.#2pt    
 \hbox{\vrule width.#2pt height#1pt \kern#1pt  
 \vrule width.#2pt}\hrule height.#2pt}}}}

\catcode`\@=12

\begin{document}

%%%%% number equations by section %%%%%%%%  
\makeatletter \@addtoreset{equation}{section} \makeatother  
\renewcommand{\theequation}{\thesection.\arabic{equation}}  

\renewcommand\baselinestretch{1.25}
\setlength{\paperheight}{11in}
\setlength{\paperwidth}{9.5in}
\setlength{\textwidth}{\paperwidth-2.4in}     \hoffset= -.3in   % +1in from printer
\setlength{\textheight}{\paperheight-2.4in}   \topmargin= -.6in % +1in from printer

\begin{titlepage}  
  
%\version\versionno  
  
%  
  
\vskip 1cm  
  
  \leftline{\tt hep-th/0512031}
\rightline{\small{\tt MCTP-05-100}}  
  
%\rightline{\small{\tt cpht-03-}}  

\vskip 3cm  
  
 \centerline{\bf \Large 
QCD recursion relations from the largest time equation }  
\vskip 1cm  
{\large }  
\vskip 1cm 

\centerline{\large  Diana Vaman, York-Peng Yao  }
  
\vskip .2cm   
\centerline{\it Michigan Center for Theoretical  
Physics}  
\centerline{ \it Randall Laboratory of Physics, The University of  
Michigan}  
\centerline{\it Ann Arbor, MI 48109-1120}  
  
\vspace{1cm}  
  
\begin{abstract}  
We show how by reassembling the tree level gluon Feynman diagrams in a 
convenient gauge, space-cone, we can explicitly derive the BCFW recursion 
relations.
Moreover, the proof of the gluon recursion relations hinges on an identity
in momentum space which we show to be nothing but the Fourier transform of 
the largest time equation. Our approach lends itself to natural generalizations
to include  massive scalars and even fermions.  
\end{abstract}  
  
\end{titlepage}

%\end{center}  
  
%\noindent  

%%%%%%%%%%%%%%%%%%%%%%%%%%%%%%%%%%%%%%%%%%%%%%%%%%%%%%%%%%%
\section{Introduction}
\indent 

QCD calculations are notoriously tedious if one is to follow the usual 
Feynman-Dyson expansion in some commonly used, such as Feynman, 
gauge.  Over the past few years, great strides have been made to simplify 
such endeavors.  The results for the complete amplitudes at the tree or one-
loop level can be quite compact.  

Following Witten's proposal for a description of perturbative 
Yang-Mills gauge theory as a string theory on twistor space
\cite{Witten:2003nn}, and subsequent proposal for an alternative to the 
usual Feynman diagrams in terms of the so-called maximally helicity
violating (MHV) vertices \cite{Cachazo:2004kj},  
a new set of methods was available for the computation of QCD amplitudes.
The latest advance in the form of recursion relations
\cite{Britto:2004ap,Britto:2005fq},  in conjunction with 
the attendant rules for their construction, is particularly appealing.  
It is quite
obvious from the flavor of such an approach that it bears on the cutting rules
in field theory.  In fact, some work at the one-loop level under the 
heading of 
cut-constructibility clearly points to the same origin \cite{Bern:1994cg}.  
These brief remarks
certainly call for the possibility to develop the subject further in the 
context 
of quantum field theory and it is our intention to do so in this short article.

As is well-known, unitarity of the S-matrix and the feasibility  of an ordering of a sequence 
of space time points are intimately related.  Indeed, the ordering need not be with
respect to time, as is conventionally done.  All that is essential in a perturbation 
series is that one must be able to separate the positive frequency and the negative frequency
components in a propagator according to the signature of a certain
linear combination of components $\Delta x$ of the four vector between the 
two space-time 
points.  For our purpose,  a component $\eta \cdot \Delta x$ of the light-cone variables 
will  be a convenient start, where $\eta$ is a light-like vector.  We shall  rely 
on the existence of tubes of analyticity to continue such variables into the space
cone, in order to incorporate a gauge condition for QCD.  The resulting ordering 
is the equivalent of the largest time equations.  We shall show that it is a consequence 
of these equations, when transcribed into momentum space, which give rise to a 
set of recursion relations.  The outcome, 
for QCD in particular, is that one factorizes
a physical amplitude into products of physical amplitudes, with some momenta
shifted but still on-shell.  This is the content of the BCFW
recursion relations \cite{Britto:2004ap,Britto:2005fq}:
$${\cal A}(P,\{ P_i\}, Q,\{ Q_j\})=\sum_{i,j} {\cal A}_L(\hat P,
\{P_i\})\frac{1}{(P+\sum_i P_i)^2}{\cal A}_R(\hat Q,\{Q_j\})\,,
$$ 
where ${\cal A}_L, {\cal A}_R$ are lower n-point functions obtained
by isolating two reference gluons with shifted momenta, $\hat P=P-z\eta$,
$\hat Q=Q+z\eta$  with $\eta^2=\eta\cdot P=\eta\cdot Q=0$, 
on the two sides of the cut. The shifting is necessary in order to 
preserve energy-momentum conservation.  
We would like to take this opportunity to point 
out that in so far as factorization is concerned, the masses of the internal 
propagators have no bearing.  However, the demand that the shifted momenta,
which will be called reference momenta, should be on-shell will force these external 
momenta to be light-like.  

We now turn to the important step of gauge fixing.  
In order to facilitate natural 
cancellation of terms at every level of a QCD calculation, 
the gauge that is most 
convenient for us is the space-cone gauge \cite{Chalmers:1998jb}.  
Here, the dependent degrees of 
freedom are completely eliminated and only two helicity components are left 
in the Lagrangian.  To accomplish that, the now complexified null vector  
$\eta $ is called upon.  Loosely speaking, it is a direct product of two spinors.  
They are used on the one hand as the 
reference spinors  separately for the two helicity components of the gauge field.  On the 
other, they will be identified with two of the external momenta in a process.
A further advantage of this gauge is that when we shift the momenta to obtain
recursion relations, the dependence on momenta of the vertices will not be 
affected.  Thereupon the factorization of the amplitudes is the same 
as that in a scalar theory.  It is this special attribute which makes the program 
manageable.

The plan of this article is as follows.  In the next section, we write down the
QCD Lagrangian in the space-cone gauge, where the auxiliary fields are 
eliminated.  The number of diagrams contributing to a process will be 
drastically reduced.  It is seen that the relevant propagator is basically 
scalar and that the vertices have good behavior under the kind of momentum
shifts we shall make.  This propagator will be decomposed into positive and 
negative frequencies in Section 3 according to light cone ordering.  Sequencing 
space-time points in this ordering will be followed in Section 4 and the largest 
time equation will be summarized.   

To familiarize the reader with space-cone calculations and the underlying 
mechanism due to momentum shifting for factorization, we give several 
simple examples in Section 5.  We want to emphasize that we obtain the 
recursion relations not only because the products of propagators 
satisfy certain algebraic 
identities, but just as  importantly because the vertices also respect the 
kinematics in the shifted momenta.  Furthermore, polarization factors have 
to be provided for the cut lines, so that the cut graphs are indeed physical 
amplitudes.  The space-cone gauge fulfills all these demands.

In Section 6, we show that the propagator identity  we used for five gluon 
amplitudes is a progeny of the largest time equation given earlier.  It is 
this deep connection that we should be able to push the factorization 
program much further into the loop level.

Section 7 is used to generalize the propagator identity to any number of 
space-time vertices which connect the flow of the two reference vectors.  
We shall
show in fact that the propagators can have any masses, leading
to a field theoretical proof of the recursion relations for massive 
charged scalars.  At the same time this observation
of course opens a new vista to include massive quarks, which we discuss 
in Section 8. 

%%%%%%%%%%%%%%%%%%%%%%%%%%%%%%%%%%%%%%%%%%%%%%%%%%%%%%%%%%
\section{The space-cone gauge fixed Yang-Mills action}
\indent

Consider the four-dimensional Yang-Mills gauge theory, with the Lagrangian
\be
L=-\frac 18 Tr(\partial^a A^b -\partial^b A^a +i  [A^a, A^b])^2\,.
\ee
Following \cite{Chalmers:1998jb}, we decompose the four-dimensional vector indices in 
a light-cone basis
\be
A^a=(a,\bar a, a^+, a^-)\,,
\ee
and we define the inner-product of two vectors by
\be
A\cdot B=a\bar b+\bar a b- a^+ b^--a^- b^+\,.
\ee
Equivalently, we can choose to decompose a four-dimensional vector into
bispinors:
\be
A^{\alpha\dot\beta}=\frac{1}{\sqrt 2}
A^a(\sigma_a)^{\alpha\dot\beta}=\begin{pmatrix}a^+&\bar a\cr 
a&a^-
\end{pmatrix}\,.
\ee
The indices are raised and lowered using the northwest r\^ule with the
matrix
\be
C_{\alpha\beta}=C_{\dot\alpha\dot\beta}=-C^{\alpha\beta}=-
C^{\dot\alpha\dot\beta}=\begin{pmatrix}
0&-i\cr i &0\end{pmatrix}\,.
\ee

Null (light-like) vectors can be decomposed into a product of two commuting 
spinors (twistors):
\be
P^{\alpha\dot\beta}=p^\alpha p^{\dot \beta}\,.
\ee
Moreover, we can use the twistors to define a basis on the space of 
four-vectors
\be
p^\alpha\equiv \langle p|, p_\alpha\equiv |p\rangle, 
p^{\dot\alpha}=[p|, p_{\dot\alpha}=|p]\,,
\ee
such that
\be
P =p^+\,|+\rangle [+| + p^-\, |-\rangle [-| + p\, |-\rangle [+| +\bar p\,
|+\rangle [-| \,.
\ee
With the normalization
\be
\langle + -\rangle = [-+]=1
\ee
it follows that the components of a null four-vector $P=|p\rangle [p|$  
onto the twistor basis are given by
\be
p^+=\langle p-\rangle[-p], p^-=\langle +p\rangle [p+], p=\langle +p\rangle
[-p], \bar p=\langle p -\rangle [p+]\,.
\ee
As shown by \cite{Chalmers:1998jb}, in a twistor formulation of the gauge 
theory a powerful simplification is 
achieved in the Feynman diagramatics by choosing the space-cone gauge:
\be
a=0\,,
\ee
followed by the elimination of the ``auxiliary'' component 
$\bar a$ from its equation of motion.
The gauge fixed Lagrangian has now only two scalar degrees of freedom
\be\label{lgf}
{\cal L}=Tr\bigg[\frac 12 a^+\Box a^-
-i\bigg(\frac{\partial^-}{\partial}a^+\bigg)
[a^+,\partial a^-]-i\bigg(\frac{\partial^+}{\partial}a^-\bigg)[a^-\partial a^+]
+[a^+,\partial a^-]+[a^+,\partial a^-]\frac{1}{\partial^2}[a^-,\partial a^+]
\bigg]\,.
\ee
Choosing the space-cone gauge amounts to selecting two of the external momenta
to be the reference null vectors for defining a twistor basis: $|+\rangle [+|,|-\rangle [-|$, such that the space-cone gauge fixing is equivalent to
$N\cdot A=0$, where the null vector $N$ is equal to $|+\rangle [-|$.
The other ingredient which is needed in converting the essentially scalar
Feynman diagrams arising from the gauge fixed Lagrangian (\ref{lgf})
into definite helicity gluon Feynman diagrams is inserting external line 
factors
\be
\epsilon^+=\frac{[-p]}{\langle + p\rangle}, \epsilon^- = \frac{\langle + p
\rangle}{[-p]}
\ee 
for the positive, respectively negative helicity external gluons.
The helicities of the internal lines/virtual gluons are accounted for by the
the scalar Lagrangian: a $``+-''$ helicity internal line corresponds to a 
$a^+ a^-$ propagator, and vice versa.

%%%%%%%%%%%%%%%%%%%%%%%%%%%%%%%%%%%%%%%%%%%%%%%%%%%%%
\section{The propagator}
\indent 

For later purposes we explicitly construct a 
representation of the Feynman propagator, wherein a light-like 
four-vector is introduced as a parameter.  Thus for a scalar field\footnote{
Recall that in the 
gauge-fixed Yang-Mills action the gauge field was reduced to 
two propagating scalar degrees of freedom. So the scalar propagator 
constructed in this section is relevant also for gauge bosons.} one writes
\bea
\Delta(x-y)&=&{1\over i} \int {d^4L\over (2\pi)^4 }
{1\over L^2-i\epsilon}e^{iL\cdot (x-y)}\label{Fp}\,.
\eea
We will show that the propagator can be equally well represented as
\bea
\Delta(x-y)
& =& {1\over i} \int {d^4p\over (2\pi)^4} \int  {dz  \over z-i\epsilon } 
\theta (p^+) \delta (p^2) e^{i(p-z\eta)\cdot (x-y)}\nonumber\\
&-&{1\over i} \int {d^4p\over (2\pi)^4} \int  {dz  \over z+i\epsilon } 
\theta (-p^+) \delta (p^2) e^{i(p-z\eta)\cdot (x-y)}\,,
\label{prop1}
\eea
with $\eta$ an arbitrary null vector.
Appropriating the following common notations to the 
light-cone frame context:  
\bea
&&\!\!\!\!\!\!\!\!\!
\delta^+(p^2)=\delta(p^2)\theta(p^+),\qquad \!\!\delta^-(p^2)=
\delta(p^2)\theta(-p^+)\\
&&\!\!\!\!\!\!\!\!\!
\Delta^+(x-y)=\int{d^4p\over (2\pi)^4}\delta^+(p^2)e^{ip\cdot (x-y)},\qquad
\!\!\Delta^-(x-y)=\int{d^4p\over (2\pi)^4}\delta^-(p^2)e^{ip\cdot (x-y)}\,,
\nonumber\\
\eea
we can rewrite the position space propagator (\ref{prop1}) using a 
light-like ordering as\footnote{A more familiar form of the 
equation (\ref{causality}), using a temporal ordering, 
is $\Delta(x-y)=\theta((x-y)^0)
\Delta^+(x-y)+\theta(-(x-y)^0)\Delta^-(x-y)$, where this time $
\Delta^\pm(x-y)=\int \frac{d^4 p}{(2\pi)^4}\delta(p^2) \theta(\pm p^0) e^{ip(x-y)}
$.}
\bea
\Delta(x-y)
& =& \theta((x-y)^+)\Delta^+(x-y)+\theta(-(x-y)^+)\Delta^-(x-y)
\label{causality}\,.
\eea
This can be seen, by first performing the contour integration over $z$
\footnote{Our conventions for the metric are mostly plus, and we 
define $x^\pm=\frac {x^0\pm x^1}{\sqrt{2}}$.}
in (\ref{prop1})
\bea
\int{dz  \over z-i\epsilon } e^{-iz\eta\cdot(x-y)}=2\pi i\theta((x-y)^+)
\nonumber\\
\int{dz  \over z+i\epsilon } e^{-iz\eta\cdot(x-y)}=-2\pi i\theta(-(x-y)^+)\,,
\eea
under the assumption that the null vector $\eta$ is equal to 
$(\eta^-, 0,\dots,0)$ and $\eta^->0$, followed by an integration over $p^-$.
We end up with
\bea
\!\!\Delta(x-y)\!\!\!&\!\!\!=\!\!\!&\!\!\!\int \!{dp^+\over 2\pi}\!\int\! {d\hat p^i\over(2\pi)^2}
\bigg(\frac{\theta(p^+)\theta((x-y)^+)}{|p^+|}+
\frac{\theta(-p^+)\theta(-(x-y)^+)}{|p^+|}
\bigg)e^{ip\cdot(x-y)}\nonumber\\&&\!\!\!.
\eea
The latter expression can be reproduced starting from the usual Feynman
propagator (\ref{Fp}), going to a light-cone frame and contour integrating 
over $L^-$.

Note that the previous discussion can be extended to include massive particles.
The minor extension involves replacing $\delta^\pm(p^2)$ by 
$\delta^\pm(p^2+m^2)$, while $\eta$ remains a null vector. 

Finally, one can easily show that the propagator as defined in 
(\ref{prop1}) obeys the Klein-Gordon equation
\bea 
(-\partial ^2 +m^2)\Delta(x-y)={1\over i} \delta ^4(x-y)\,.
\eea

%%%%%%%%%%%%%%%%%%%%%%%%%%%%%%%%%%%%%%%%%%%%%%%%%%%%%%%%%%%%%%
\section{The causality (``largest time'') equations}
\indent

As stated in the Introduction, we shall show that the recursion relations 
are rooted in the largest time equation. To this end, 
we briefly revisit here the causality equations as derived by 
Veltman\cite{Veltman:1963th}, 
but appropriately rewriting them in a light-cone frame.

First, we introduce the following set of rules:

-duplicate the Feynman diagram $2^N$ times, for $N$ vertices, by adding
circles around vertices in all possible ways;

-each vertex can be circled or not; a circled vertex will bring a factor
of $i$, and an uncircled vertex will bring a factor of $(-i)$;

-the propagator between two uncircled vertices is $\Delta(x-y)$, 
while the propagator between two circled vertices is the complex conjugate
$\Delta^*(x-y)$;

-the propagator between a circled $x_k$ and an uncircled $x_l$ is 
$\Delta^+(x_k-x_l)$, while the propagator between an uncircled $x_k$ and 
a circled $x_l$ is $\Delta^-(x_k-x_l)$.

Clearly, the uncircled Feynman diagram is the usual one, while the
fully circled diagram corresponds to its complex conjugate.

The largest time equation states that the sum of all $2^N$ circled Feynman
diagram vanishes:
\be
F(x_i)+F^*(x_i)+{\bf F}(x_i)=0\label{causeqn}\,,
\ee 
where $F(x_i)$ stands for the usual Feynman diagram, $F^*(x_i)$ is its
complex conjugate, and ${\bf F}(x_i)$ is the sum of $2^{N}-2$ diagrams in 
which at least one vertex is circled and at least one is uncircled.

Other causality equations can be obtained by singling out 2 vertices, 
$x_k$ and $x_l$. Let us assume $x_k^+<x_l^+$. Then, one has
\be
\theta((x_l-x_k)^+)(F(x_i)+{\bf F}(k,x_i))=0\label{causeqn1}\,,
\ee
where ${\bf F}(k,x_i)$ is the sum of all diagrams with $k$ uncircled, 
but at least one other vertex circled. Similarly, one has
\be
\theta((x_k-x_l)^+)(F(x_i)+{\bf F}(l,x_i))=0\label{causeqn2}\,,
\ee
By adding these two equations one finds
\be
F(x_i)=-{\bf F}(k,l,x_i)-\theta((x_l-x_k)^+){\bf F}(k,{\bf l},x_i)-
\theta((x_k-x_l)^+){\bf F}({\bf k},l,x_i))\label{caus}\,,
\ee
where ${\bf F}(k,l,x_i)$ is the sum of all diagrams with neither $k,l$ circled,
but at least one other vertex circled, ${\bf F}(k,{\bf l}, x_i)$ is the sum 
of all amplitudes with $k$ uncircled , but $l$ circled and finally, 
${\bf F}({\bf k},l,x_i))$ has $k$ circled and $l$ uncircled.

By taking the Fourier transform of the real part of the position
space Feynman diagrams $\frac 1i\int dp (F(x_i)+F^*(x_i))$ one obtains 
the imaginary part of the momentum space diagram.
The cut graphs (Cutkosky rule) are derived by the use of (\ref{causeqn}).
The momentum space propagators between two vertices 
with both vertices uncircled is 
(\ref{Fp}) $-i/(p^2-i\epsilon)$, the complex conjugate expression if the two
vertices are circled, and $2\pi\delta^+(p^2)$ if the momentum flows between
an uncircled and a circled vertex.

%%%%%%%%%%%%%%%%%%%%%%%%%%%%%%%%%%%%%%%%%%%%%%%%%%%%%%%%%%%%%%%%%%%
\section{Reassembling Feynman diagrams into BCFW \\recursion relations}
\indent

The Feynman diagrams, as advertised, are the ones following from the 
space-cone gauge-fixed Lagrangian (\ref{lgf}).
To set the stage for the recursion relations involving arbitrary
tree level gluon n-point functions, we begin our investigation 
with the lowest ones. 

The 3-point function is the same as the 3-point vertex up to multiplication 
by the external lines polarization vectors. Otherwise, the 3-point function
has only one peculiarity: to define it, one cannot select the 2 reference 
gluons out of the 3 external gluons. Take for instance $(123)=(++-)$, with
1 selected as reference and an arbitrary null vector $4=\ket{4}\sbra{4}$.
The 3-point function is given by
\bea
(++-)=\epsilon_2^+ \epsilon_3^- k_3=
\bigg(\frac{[12]}{\langle 42\rangle}\frac{\langle43\rangle}{[13]}\langle 42
\rangle [12]\bigg)\frac{1}{\langle 14\rangle }\,,
\eea
where the factor $1/{\langle 14\rangle }$ is inserted as a matter 
of normalization of the angular and square brackets. Using that
\bea
\langle 43\rangle [23]=\langle 41\rangle [12]\,,
\eea
the 3-point function acquires the standard googly-MHV expression
\bea
(++-)=\frac{[12]^3}{[23][31]}\,.
\eea

We now proceed to evaluate the 4-point functions. This is our first demonstration of factorization, which will be cast into a BCFW recursion. Let us consider 
$(1234)=(+--+)$, with 1 and 2 selected the reference gluons
\bea
&&\ket{1}=\ket{-},\qquad \sket{1}=\sket{-}\\
&&\ket{2}=\ket{+}, \qquad \sket{2}=\sket{+}\,.
\eea
In this basis, the polarization vectors are 
\bea
&&
\epsilon_p^+=\frac{[-p]}{\langle + p\rangle }=\frac{[1p]}{\langle 2p\rangle}\\
&&
\epsilon_p^-=\frac{\langle +p \rangle}{[-p]}=\frac{\langle 2p\rangle}{[1p]}\,,
\eea
and the 4 components of the momenta decomposed in the bispinor basis read 
\bea
&&p^+=[-p]\langle p-\rangle=[1p]\langle p1\rangle\\
&&p^-=[+p]\langle p+\rangle=[2p]\langle p2\rangle\\
&&p=-[-p] \langle p+\rangle=-[1p]\langle p 2\rangle\\
&&\bar p=-[+p]\langle p-\rangle=-[2p]]\langle p1\rangle\,.
\eea
Notice that the only non-vanishing components of the reference gluon momenta 
are
\bea
p_1^-=1, \qquad p_2^+ =1\,.
\eea
We now pose the question what happens with the Feynman diagrams 
if for later purposes we decompose  
the $\eta$-shifted reference gluon momenta, with the choice
\be
\eta=\ket{2}\sbra{1}\,,
\ee
in the 
{\it{same twisor basis}}? Concretely, we define the $\eta$-shifted reference
gluon momenta to be
\be
\widehat{P_1}=P_1+z\eta, \widehat{P_2}=
P_2-z\eta,\label{shift}\,,
\ee 
or, in components,
\bea
&&\hat p_1^-=1, \qquad \bar{\hat p}_1=-z[21]\langle21\rangle\\
&&\hat p_2^+=1,\qquad \bar{\hat p}_2=z[21]\langle 21\rangle\,.
\eea 
Thus the only changes to the reference momenta enter through the component 
$\bar p$. This is particularly important, since the vertices in the space-cone 
gauge turn out to be independent of $\bar p$, as it can be seen by inspecting
the Lagrangian (\ref{lgf}).

{\it We conclude that any shift of the external momenta 
as in (\ref{shift}) is of no consequence
for the vertex-dependence of any Feynman diagram, and leaves an imprint only
over the internal line propagators.}

The 4-point function in the space-cone gauge 
is given by a single Feynman diagram
\\ 
\begin{figure}[h]
\includegraphics[width=6in]{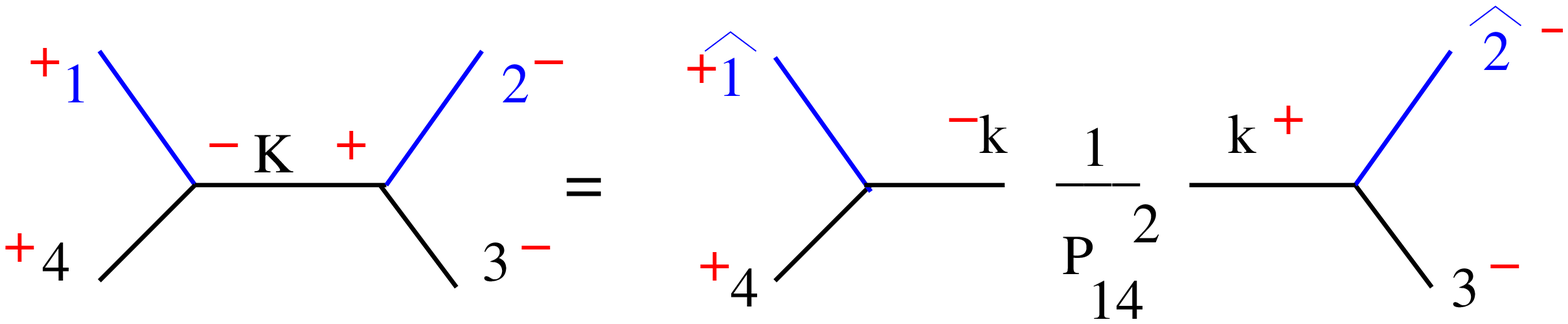}
\caption{Factorization of the 4-point function}
\end{figure}
\\
which is represented by the left hand side of Fig.1.
With the choices already made in terms of reference momenta\footnote{
The factor of $1/2$ in the propagator is due to a peculiar normalization
of \cite{Chalmers:1998jb}.}
\bea
(1234)=(+--+)&=&\epsilon_3^-\epsilon_4^+ p_K^2 \frac{1}{\frac 12 
P_{14}^2}\frac{1}{[12]
\langle 12\rangle}\\
&=&\frac{\langle 23\rangle}{[13]}\frac{[14]}{\langle 24 \rangle}
[13]\langle 32\rangle [14]\langle 42 \rangle \frac{1}{[14]\langle 41\rangle}\frac{1}{[12]
\langle 12\rangle}\\
&=&\frac{\langle 23\rangle^2 [14]}{\langle 41\rangle\langle 12\rangle [12]}\,.
\eea
Using that $[12]\langle 23\rangle=-[14]\langle 43\rangle$, which follows from
momentum conservation one recovers the simple MHV expression of the 
4-point function
\bea
(1234)=(+--+)=\frac{\langle 23\rangle^3}{\langle 12\rangle\langle 34\rangle
\langle 41 \rangle}\,.
\eea

On the other hand, we may choose to split the 4-point function into
lower on-shell amplitudes, 3-point functions, as indicated on the right 
hand side of Fig.1.
The internal leg $K$ has been put on-shell by a shift of the two
external reference gluons as in (\ref{shift}), where
\bea
z&=&-\frac{P_{14}^2}{2\eta \cdot P_{14}}
=-\frac{\langle 14\rangle [14]}{\langle 24\rangle [14]}=
-\frac{\langle 14\rangle}{\langle 24\rangle}\,.
\eea

Notice that because there is no difference between the 3-point function and 
3-point vertex, other than the multiplication by external line polarizations,
 and because we have essentially a scalar field theory, one can insert freely
factors $\epsilon_k^+\epsilon_k^-$ (=1). Recalling that as argued before, the
shift (\ref{shift}) in the external momenta does not modify the vertices,  
we see that the factorization into 3-point functions according to Fig.1. 
is trivially realized. Furthermore, if we follow the same steps we made earlier
in converting the 3-point functions into MHV vertices, we arrive at the BCFW
result.

The 5-point function is the first non-trivial example in which we invoke an identity rooted in the largest time equation. 
To begin with, it is given by the sum of three Feynman diagrams
which are represented on the left hand side of Fig.2. The right hand side
contains terms corresponding to having cut the 5-point function in all possible
ways such that the reference gluons 1,5 are on opposite sides of the 
cut. Moreover, we shift the reference external gluons such that the cut line
is on-shell. Thus the cut diagrams become on-shell amplitudes. Moreover,
the cut diagrams are multiplied  by the propagator of the line
that was cut.  
\\
\begin{figure}[h]
\includegraphics[width=6in]{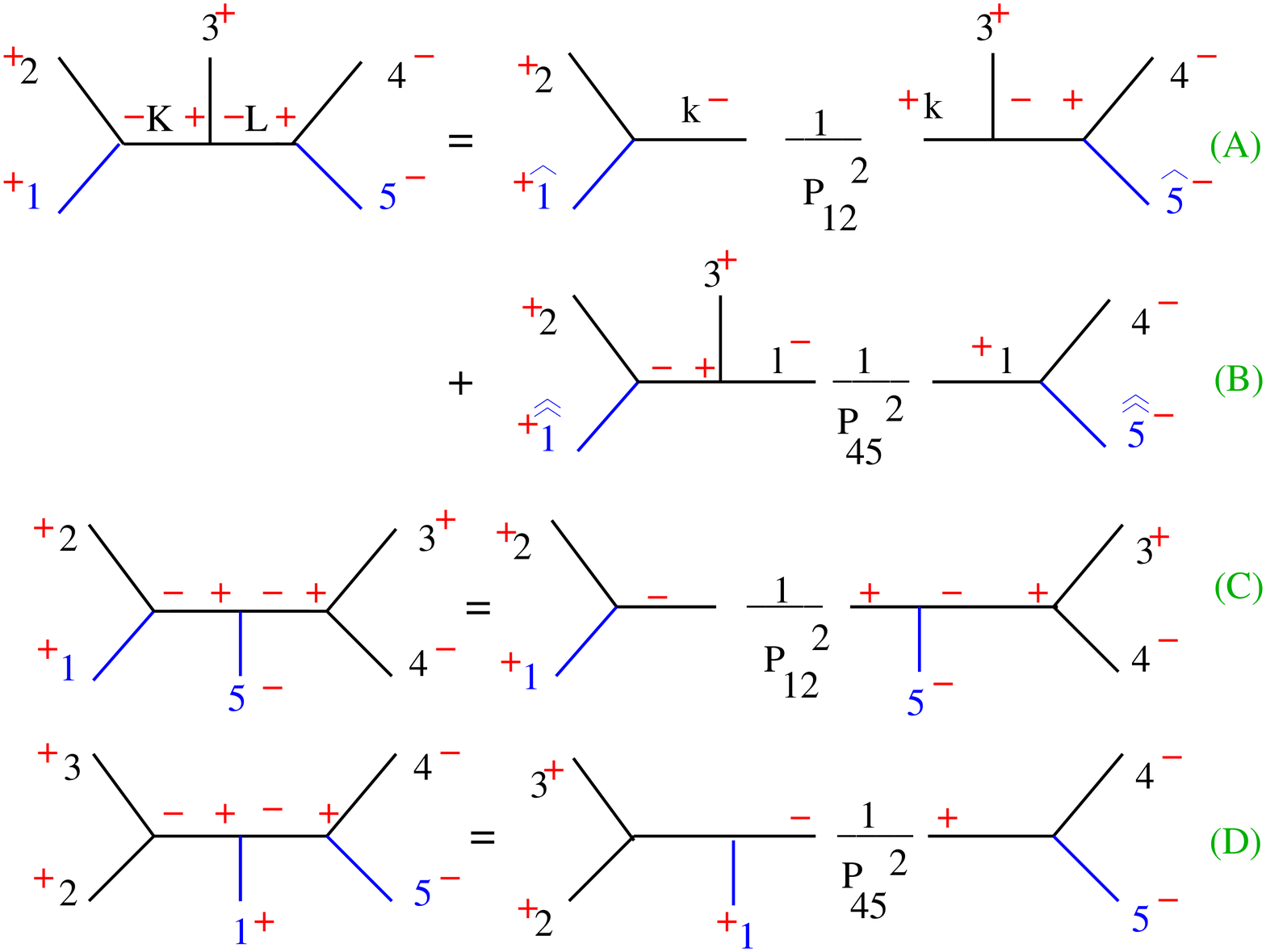}
\caption{Factorization of the 5-point function}
\end{figure}
\\
We will establish an identity relating the 5-point function, as computed
from Feynman diagrams, to the tree amplitudes of 3 and 4-point functions
as indicated by Fig.2. This, of course, is nothing but the statement of the
BCFW recursion relation, applied to this particular 5-point function.

In particular we will show that the top Feynman diagram is the sum of two such
cut diagrams (A+B), and that the middle and bottom Feynman diagrams are equal,
respectively, with C and D. 
The equality of the two sides of the last two diagrams 
is obvious from the fact that, as argued before, the 
vertices on the left and right hand side of Fig.2 are the same, irrespective 
of having shifted the external reference momenta in order to put the cut line
on shell. Moreover, the propagators  on the right and left 
hand side of the bottom two diagrams coincide as well.

The attentive reader could observe that B+D add up to zero. 
After giving an obvious
common factor, B+D add up to the $(+++-)$ 4-point function. The reason why
we have to consider two Feynman diagrams to recover the 4-point function, 
as opposed to our previous calculation,  
is that we
have selected only one of the four external gluon momenta as reference vector.
This means that we have to consider both the $s$ and $t$ channel. Nonetheless,
the sum of these two channels is zero, as it corresponds to having all but one
external gluons of the same helicity. Thus indeed, the BCFW recursion relation
amounts to including only the terms A and C, corresponding to a factorization
of the $(+++--)$ amplitude into $(++-)(++--)$ amplitudes.

In the derivation of the recursion relation directly from 
Feynman diagrams, it is useful to keep all possible terms that arise from
cutting an internal line of all Feynman diagrams, such that
the reference external gluons are on opposite sides of the cut. 
To show that the top Feynman diagram equals A+B, once we factored out the
vertices (which are the same on the left and right side of Fig.2), amounts
to proving the following identity between propagators:
\bea
\frac{1}{P_{12}^2}\frac{1}{P_{45}^2}=\frac{1}{P_{12}^2}\frac{1}{P_{4\hat 5}^2}
+\frac{1}{P_{\hat{\hat 1}2}^2}\frac{1}{P_{45}^2}\,,
\label{id5pt}
\eea 
where we defined the shifted reference gluon external momenta
\bea
&&P_{\hat 1}=P_1+\hat z\eta, \qquad P_{\hat 5}=P_5-\hat z\eta\nonumber\\
&&P_{\hat{\hat 1}}=P_1+\hat{\hat z}\eta, \qquad 
P_{\hat {\hat 5}}=P_5-\hat{\hat z}
\eta\,.
\label{5gs}
\eea
$\hat z, \hat{\hat z}$ are such that we put the internal lines $K, L$,
respectively, on-shell
\bea
&&\hat z=-\frac{P_{12}^2}{2 \eta\cdot P_{12}}=-\frac{\langle 12\rangle}{\langle
52\rangle}\label{hz}\\
&&\hat{\hat z}=\frac{P_{45}^2}{2\eta\cdot P_{45}}=
\frac{\langle 45\rangle}{\langle 15\rangle}\label{hhz}\,,
\eea
and where the null vector $\eta$ is defined, as before, with respect to 
the reference gluons 
\bea
\eta=\ket{+}\sbra{-}\equiv \ket{5}\sbra{1}\label{eta5}\,.
\eea
Despite the fact that we can prove (\ref{id5pt}) by going into the bispinor
basis, we find it much simpler and easily allowing for generalizations to
stay in momentum space.
Using that
\bea
P_{12}^2=-2\hat z\eta\cdot P_{12}, \qquad 
P_{45}^2=2\hat{\hat z}\eta\cdot P_{45}\\
P_{\hat{\hat 1}2}^2=2(\hat{\hat z}-\hat z)\eta\cdot P_{12},
\qquad P_{4\hat 5}^2=2(\hat{\hat z}-\hat z)\eta\cdot P_{45} \,,
\eea  
(\ref{id5pt}) becomes a trivial algebraic identity
\bea
\frac{1}{\hat z\hat{\hat z}}=\frac{1}{(\hat {\hat z}-\hat z)\hat{ z}}
-\frac{1}{(\hat{\hat z}-\hat z)\hat {\hat z}}\,.
\eea
This completes the proof of the BCFW recursion relation from Feynman diagrams
for the 5-point function and highlights the pattern that we will encounter
for an arbitrary n-point function.
%%%%%%%%%%%%%%%%%%%%%%%%%%%%%%%%%%%%%%%%%%%%%%%%%%%%%%%%%%%%%%%%%%%%
\section{The recursion relations and the largest time equation} 

\indent 

There is yet another way to address the identity (\ref{id5pt}) by 
rewriting the propagators in position space and next recognizing the 
Fourier transform of the (i.e. momentum space) largest  
time equation (\ref{caus}). Reinstating the usual $i\epsilon$ prescription
in the momentum space propagators, and multiplying (\ref{id5pt}) with 
the total momentum conservation $\delta$-function, the right-hand-side
of (\ref{id5pt}) becomes:
\bea
\frac{\delta(P_1+\dots+P_5)}{P_{12}^2}\frac{1}{P_{45}^2}
\!&=&\!\int \!x_1 dx_2 dx_3\! \int \!\frac{dK}{(2\pi)^4}\! 
\int \!\frac{dL}{(2\pi)^4} e^{i(P_1+P_2+K)x_1}
e^{i(P_3-K+L)x_2} e^{i(P_4+P_5-L)x_3}\nonumber\\
&&\frac{1}{K^2-i\epsilon}\frac{1}{L^2-i\epsilon}
\nonumber\label{s1}\\
&\!=&\!i^2\int dx_1 dx_2 dx_3 \Delta(x_1-x_2)\Delta(x_2-x_3)
e^{i(P_1+P_2)x_1+iP_3 x_2+i(P_4+P_5)x_3}.
\eea
The shifted propagators which appear on the left-hand-side of
(\ref{id5pt}) can be cast into 
\bea 
\!\!\!\!\!\!\!\!\!\!\!\!&&\!\!\!\!\!\!\!\!\!\!\!\!\!
\frac{\delta(P_1+\dots+P_5)}{P_{12}^2}\frac{1}{P_{4\hat 5}^2}\!=\!
\int \!dx_1 dx_2 dx_3 \!\int\! \frac{dk}{(2\pi)^4} 
\!\int\! dz \!\int \!
\frac{dL}{(2\pi)^4} e^{i(P_1+P_2+k)x_1} e^{i(P_3-k+L+z\eta)x_2} 
e^{i(P_4+P_5-L-z\eta)x_3}\nonumber\\
&&\;\;\;\;\;\;\;\;\;\;\;\;\;\;\;\;\;\;\;\;\;\;\;
\;\;\;\;\;\;\frac{1}{L^2-i\epsilon}\bigg(
\frac{\delta^-((k-z\eta)^2)}{z-i\epsilon}-\frac{\delta^+((k-z\eta)^2)}
{z+i\epsilon}\bigg)\nonumber\\
&&~~~~~~~~~~~~~~~~~~~~=
i^2\int dx_1 dx_2 dx_3 e^{i(P_1+P_2+)x_1+iP_3x_2+i(P_4+P_5)x_3}
\nonumber\\&&~~~~~~~~~~~~~~~~~~~~
\bigg(\theta((x_1-x_3)^+)\Delta^+(x_1-x_2)+
\theta((x_3-x_1)^+)\Delta^-(x_1-x_2)\bigg) \Delta(x_2-x_3),
\label{s2}
\eea
using the same z-parametrization which we have introduced in Section 3.
For concreteness we have chosen, as before, $\eta=(\eta^-, 0,\vec 0)$ 
with $\eta^->0$.
Clearly, by integrating out first $x_1,x_2,x_3$ and using the delta-function
to integrate over $k, L$, followed by a $z$-integration using the remaining
delta-function $\delta((P_1+P_2-z\eta)^2)$, we recover the left-hand-side
of (\ref{s2}).
On the other hand, if we choose to shift the integration variable
from $k$ to $k-z\eta$, and we next integrate over $z$, then we find
the result given in the last line of (\ref{s2}).

Similarly, the other term on the left-hand-side of (\ref{id5pt}) can be 
written as
\bea 
\!\!\!\!\!\!\!\!\!\!\!\!&&\!\!\!\!\!\!\!\!\!\!\!\!\!
\frac{\delta(P_1+\dots+P_5)}{P_{1\hat{\hat 2}}^2}\frac{1}{P_{4 5}^2}\!=\!
\int \!dx_1 dx_2 dx_3 \!\int \!\frac{dK}{(2\pi)^4} \!
\int \!\frac{dl}{(2\pi)^4}\!\int \!dz\; e^{i(P_1+P_2+K+z\eta)x_1} e^{i(P_3-K-z\eta+l)
x_2} e^{i(P_4+P_5-l)x_3}\nonumber\\
&&\;\;\;\;\;\;\;\;\;\;\;\;\;\;\;\;\;\;\;\;\;\;\;
\;\;\;\;\;\;\frac{1}{K^2-i\epsilon}\bigg(
\frac{\delta^-((l-z\eta)^2)}{z-i\epsilon}-\frac{\delta^+((l-z\eta)^2)}
{z+i\epsilon}\bigg)\nonumber\\
&&~~~~~~~~~~~~~~~~~~~~=
i^2\int dx_1 dx_2 dx_3 e^{i(P_1+P_2)x_1+iP_3x_2+i(P_4+P_5)x_3}
\nonumber\\&&~~~~~~~~~~~~~~~~~~~~
\bigg(\theta((x_1-x_3)^+)\Delta^+(x_2-x_3)+
\theta((x_3-x_1)^+)\Delta^-(x_2-x_3)\bigg) \Delta(x_1-x_2)
\label{s3}.
\eea
Thus the identity (\ref{id5pt}) becomes
\bea
\int &&\!\!\!\!\!\!\!\!\!\!\!\!\!\!
dx_1 dx_2 dx_3 e^{i(P_1+P_2)x_1+iP_3x_2+i(P_4+P_5)x_3}\bigg[
\Delta(x_1-x_2)\Delta(x_2-x_3)\nonumber\\
&-&\bigg(\theta((x_1-x_3)^+)\Delta^+(x_1-x_2)+
\theta((x_3-x_1)^+)\Delta^-(x_1-x_2)\bigg) \Delta(x_2-x_3)\nonumber\\
&-&\bigg(\theta((x_1-x_3)^+)\Delta^+(x_2-x_3)+
\theta((x_3-x_1)^+)\Delta^-(x_2-x_3)\bigg) \Delta(x_1-x_2)
\bigg]=0\label{ft5pt}.
\eea

There is one more step that is needed in order to show the relationship 
between 
(\ref{id5pt}) and the largest time equation. From (\ref{caus}), with 
$x_1,x_3$ the two vertices that are singled out, we have 
\bea
\Delta(x_1-x_2)\Delta(x_2-x_3)&=&\Delta^-(x_1-x_2)\Delta^+(x_2-x_3)\nonumber\\
&+&\theta((x_1-x_3)^+)\bigg(\Delta^+(x_1-x_2)\Delta(x_2-x_3)-\Delta^*(x_1-x_2)
\Delta^+(x_2-x_3)\bigg)\nonumber\\
&+&\theta((x_3-x_1)^+)\bigg(\Delta(x_1-x_2)\Delta^-(x_2-x_3)
-\Delta^-(x_1-x_2)\Delta^*(x_2-x_3)\bigg).\nonumber\\
\label{c5pt}
\eea
To show how (\ref{ft5pt}) is related to (\ref{c5pt}) we first rearrange 
the right-hand-side of (\ref{c5pt}) using
\be
\Delta^*(x-y)=\theta((x-y)^+)\Delta^-(x-y)+\theta((y-x)^+)\Delta^+(x-y)\,,
\ee
such that it becomes equal to the Fourier transform of the 
right-hand-side of equation (\ref{ft5pt}), up to the following two extra 
terms: $\theta((x_1-x_3)^+)\Delta^+(x_1-x_2)\Delta^+(x_2-x_3)$ and $
\theta((x_3-x_1)^+)\Delta^-(x_1-x_2)\Delta^-(x_2-x_3)$.
These terms in fact are zero as the product of the three distributions 
has zero support. The easiest to see this is to evaluate the Fourier transform
\bea
\int dx_1 dx_2 dx_3 e^{i(P_1+P_2)x_1+iP_3x_2+i(P_4+P_5)x_3}
\theta((x_1-x_3)^+)\Delta^+(x_1-x_2)\Delta^+(x_2-x_3)
\eea
by rewriting the step function as a $z$-integral, followed by the integration
over $x_1,x_2,x_3$, to arrive at
\bea
\delta(P_1+\dots P_5)\int dz \frac{1}{z-i\epsilon}
\delta^+((P_1+P_2+z\eta)^2)\delta^+((P_4+P_5-z\eta)^2)\,.
\eea
It is clear that no $z$ can satisfy the simultaneously the two 
delta-function constraints. This completes the proof that the algebraic 
identity which was found by reassembling the Feynman diagrams into the BCFW
recursion relations arises from the more fundamental largest time equation.

Before closing this section, we need to add some remarks to explain
some more what we have accomplished.  The discussion above of 
the largest time equation is taken for real $\eta $, since it is only for 
such values that we know how to order a sequence of space-time 
points  in $\eta \cdot x_i$.  

Now that we have the largest time equation, let us Fourier- transform 
it into momentum space, appropriate for a physical process under 
consideration and keeping $\eta$ as a variable.  Then we obtain a 
set of shifted momenta, as in (\ref{5gs}).  We then analytically 
complexify $\eta$.  For tree level, this is certainly possible 
and justifiable, 
because the dependence on it is only in some algebraic functions of propagators.
At the loop level, we need to invoke the analysis of axiomatic 
field theorists \cite{book}, 
which states that  there are tubes of analyticity to allow this extension and 
to lead to complexified unitarity relations.  We now identify these propagators 
with the ones which we need in the space-cone gauge to carry on with the 
analysis.

%%%%%%%%%%%%%%%%%%%%%%%%%%%%%%%%%%%%%%%%%%%%%%%%%%%%%%%%%%%%%%%%%%%%

\section{The general case}
\indent 

To exploit the full generality of the problem, we derive an 
identity satisfied by the momentum space scalar propagators
working under the assumption that we deal with massive propagators, 
with arbitrary masses. 

Consider a graph at the tree level with $n'$ vertices and $m' $ 
external lines which 
are on-shell.  As our convention, we take them to be all outgoing.  
We single out two of these lines which do not land on the same vertex 
as reference vectors and call 
them $p_a$ and $p_b$.  For a tree graph, there is a unique path through some 
of the internal lines which connects $p_a$ to $p_b$.  We shall denote the 
vertex at which 
$p_a$ emanates $x_1$, and that for $p_b$ $x_n$ in their space-time labels.
The vertices in between are $x_i, \  i=2, \dots n-1.$  There are then $n$ 
vertices in this segment of the graph and therefore n-1 internal lines.  
Our consideration for the time being will be on this segment.   The internal 
lines carry momenta $q_i$, joining $x_i$ to $x_{i+1}, n-1 \ge  i \ge 1.$  
How and what other lines enter or leave these vertices need not concern us 
at this point. For each 
$q_i$, we associate a propagator
\bea
 \Delta(x_i-x_{i+1}, m_i)&=&
\int {d^4q_i\over (2\pi)^4i}{e^{iq_i\cdot (x_i-x_{i+1})}
\over q_i^2+m_i^2 -i\epsilon} \nonumber\\ 
&=&\theta (\eta \cdot (x_i-x_{i+1}))\Delta^+(x_i-x_{i+1}, m_i)
+\theta (-\eta \cdot (x_i-x_{i+1}))\Delta^-(x_i-x_{i+1}, m_i)\,,
\nonumber\\
\eea
where
\bea
\Delta^+(x_i-x_{i+1}, m_i)=\int {d^4\bar q_i\over (2\pi)^4i}e^{i\bar q_i\cdot (x_i-x_{i+1})}
\delta^+(\bar q_i^2+m_i^2),\nonumber\\
\Delta^-(x_i-x_{i+1}, m_i)=\int {d^4\bar q_i\over (2\pi)^4i}e^{i\bar q_i\cdot (x_i-x_{i+1})}
\delta^-(\bar q_i^2+m_i^2)\,,
%\nonumber\\
%\theta (\eta \cdot (x_i-x_{i+1}))={1\over 2\pi i}\int {dz\over z-i\epsilon} 
%e^{iz\eta \cdot (x_i-x_{i+1})},
\eea
%and 
%\bea
%\theta (-\eta \cdot (x_i-x_{i+1}))=-{1\over 2\pi i}\int {dz\over z+i\epsilon} 
%e^{iz\eta \cdot (x_i-x_{i+1})}.
%\eea
and where $\eta$ is a light-like vector.  
Due to momentum conservation, each $q_i$
is expressible in terms of external momenta, and in particular 
it has a component $+ p_a$, or equivalently $-p_b$.

The factorization procedure is to cut these $q_i$ successively by shifting them by $z \eta$.  The on-shell conditions 
\bea
\bar q_i^2+m_i^2=0, \ \ \bar q _i\equiv q_i-z
\eea
will give us a set of solutions, points in the complex plane, namely
\bea
z_i={q_i^2+m_i^2\over 2 \eta \cdot q_i},
\eea
for each $ n-1 \ge i \ge 1. $
More precisely stated, the factorization amounts to splicing the graph into a 
sum of products of two on-shell graphs with shifted momenta 
$ \{p_a-z_i\eta,  \dots , \bar q_i\} $ and 
$ \{ -\bar q_i , \cdots , p_b+z_i \eta \}$, where $\cdots $ stand for 
the other momenta 
in the left graph segment and similarly for those in the right graph 
segment, with 
the propagator ${1\over q_i^2+m_i^2} $ as the partition.  
The reason that $p_a$ and $p_b$ are 
shifted  is because we need to conserve the overall momenta on the 
left and the right segment separately to make them into physical amplitudes.  
We must 
demand on-shell conditions  for the shifted $p_{a,b}$ 
with the same masses, which give
$$(p_a-z_i\eta)^2+ m_a^2=0, \ \ \ \ \  (p_b+ z_i\eta)^2+ m_b^2=0,$$
or
$$p_a \cdot \eta =0, \ \ \ \ p_b \cdot \eta =0.$$
As $\eta$ is light-like, these conditions clearly do not allow $p_{a,b}$ to be 
time-like.  Therefore, the two reference vectors must also be light-like, 
$m_a=m_b=0.$ 

The identity which we want to establish is 
\bea
{1\over q_1^2+m_1^2}{1\over q_2^2+m_2^2}\cdots   {1\over q_{n-1}^2+m_{n-1}^2}
\!\!&=&\!\!{1\over q_1^2+m_1^2}{1\over (q_2-z_1\eta)^2+m_2^2}\cdots {1\over (q_{n-1}-z_1\eta)^2+m_{n-1}^2}\nonumber\\
%\cr 
\!\!&+&\!\!{1\over (q_1-z_2\eta)^2+m_1^2}{1\over q_2^2+m_2^2}\cdots {1\over (q_{n-1}-z_2\eta)^2+m_{n-1}^2}\nonumber\\
%\cr 
\!\!&+& \!\!\cdots  \cdots \cdots\nonumber\\
\!\!&+&\!\!\!\!{1\over (q_1-z_{n-1} \eta)^2+m_1^2}
\cdots {1\over (q_{n-2}-z_{n-1}\eta)^2+m_{n-2}^2} {1\over q_{n-1}^2+m_{n-1}^2}.
\label{momid}\nonumber\\
\eea

The proof of this identity is quite simple.  
For $n-1\ge i\ne j \ge 1$, we write 
\bea
(q_i- z_j\eta)^2+m_i^2=q_i^2+m_i^2-2z_j \eta \cdot q_i\,.
\eea
Then, using the on-shell conditions for the shifted internal momenta, 
which is tantamount to making cuts, we have 
\bea
\bar q_i^2+m_i^2=0 \ \to \ q_i^2+m_i^2=2z_i\eta \cdot q_i\,.
\eea
Together, they yield
\bea
(q_i- z_j\eta)^2+m_i^2=2\eta \cdot q_i (z_i-z_j)\,.
\eea
Putting these together, we see the identity holds if one can show
\bea 
{(-1)^n\over z_1 z_2 \cdots z_{n-1}}&=&{1\over z_1 (z_1-z_2) (z_1-z_3)
\cdots (z_1-z_{n-1})}\nonumber\\
%\cr 
& +&{1\over  (z_2-z_1) z_2(z_2-z_3)\cdots (z_2-z_{n-1})}\nonumber\\
%\cr
&& \cdots \cdots \cdots\nonumber\\
%\cr 
&+& {1\over  (z_{n-1}-z_1) (z_{n-1}-z_2)\cdots (z_{n-1}-z_{n-2})z_{n-1}}\,.
\label{zid}
\eea
This is so, because (\ref{zid}) is just a formula of partial fractioning, 
or it is just a statement that the integral
$$\int {dz\over z(z-z_1)(z-z_2)\cdots (z-z_{n-1})}=0$$
for a complex variable z over a contour which encloses all the poles.

Notice that eqn. (\ref{momid}) is precisely the identity needed to reassemble
a generic tree level gluon Feynman diagram into 
lower on-shell amplitudes, as shown in Section 5. 
The reason for this is that, as we argued before, {\it the vertices
which enter the Feynman diagram and the corresponding lower n-point functions
are the same}, being insensitive to the shift of the reference gluons.
Also the external line factors that have to be inserted on the cut lines
to recover the lower on-shell amplitudes cancel pairwise, i.e. 
$\epsilon^+\epsilon^-=1$.
Then, one is left to prove only an identity involving the momentum space
scalar propagators. This is the same as (\ref{momid}), with all propagators
being massless $m_i=0$. The combinatorics work out properly to reproduce the BCFW recursions.
We have also checked these points explicitly for the six gluon amplitudes. 

Furthermore, the arguments presented in Section 6, relating the momentum
space identity (\ref{momid}) to the Fourier transform of the corresponding
largest time equation (\ref{caus}), with the singled out vertices 
corresponding to those of the external reference gluons, can be easily 
carried through.

This completes our purely field theoretical proof of the BCFW recursion 
relations. In the process, we have identified the underlying principle 
behind them in the form of the largest time equation.
 
\section{Adding massive scalars and fermions}

\indent

Establishing recursion relations to include charged 
massive scalars is straightforward 
in our framework. The current interaction term $(\Phi^* \partial_\mu\Phi -
\Phi\partial_\mu\Phi^*) A^\mu$ has only $\partial^+,\partial^-$ and $\partial$
derivatives, without the dangerous $\bar\partial$ which would have been
sensitive to the shift of the external momenta, because of the
space-cone gauge ($a=0$). The quartic interaction $\Phi\Phi^* A_\mu A^\mu$
has no momentum dependence. Therefore the vertices are unchanged under the 
$z\eta$ shifts. Besides there is no external line factor for scalars.
This implies that we have met all the requirements to accommodate recursion 
relations as stressed over and again.

A more interesting and natural generalization is the 
inclusion of fermions in the tree level recursion relations.
Consider the Lagrangian of minimally coupled massive fermions
\be
{\cal L}_f=\sum_i \bar \Psi_i(i\slash\!\!\!\nabla -m_i)\Psi_i\,,
\ee
where $\nabla_\mu$ is the gauge covariant derivative and the Dirac matrices
obey the usual anticommutation relations 
$\{\gamma^\mu,\gamma^\nu\}=-2g^{\mu\nu}$.
Then the recursion relations which are formulated with the two reference gluons
connected by a path which includes a fermionic line are based on the 
following identity involving momentum space fermion propagators
\footnote{For concreteness we considered here a quark line. Otherwise, for an antiquark the signs in $\slash \!\!\!q_i-z_j
\slash \!\!\!\eta+m_i$ must be flipped to $-\slash \!\!\!q_i+z_j
\slash \!\!\!\eta+m_i$.} 
\bea
&&{1\over \slash \!\!\!q_1 +m_1}\gamma^{\mu_1}{1\over \slash \!\!\!q_2+m_2}
\gamma^{\mu_2}
\cdots  \gamma^{\mu_{n-2}} {1\over \slash \!\!\!q_{n-1}+m_{n-1}}\nonumber\\
&=&{ m_1-\slash \!\!\!q_1+z_1\slash \!\!\!\eta\over (q_1^2+m_1^2)}
\gamma^{\mu_1}
{1\over \slash \!\!\!q_2-
z_1\slash \!\!\!\eta+m_2}\gamma^{\mu_2}\cdots \gamma^{\mu_{n-2}}
{1\over \slash \!\!\!q_{n-1}-
z_1\slash \!\!\!\eta+m_{n-1}}\nonumber\\
%\cr \!\!\!\!
&+&{1\over \slash \!\!\!q_1-z_2\slash \!\!\!\eta+m_1}\gamma^{\mu_1}
{m_2 -\slash \!\!\!q_2+z_2\slash \!\!\!\eta \over(q_2^2+m_2^2)}
\gamma^{\mu_2}
\cdots \gamma^{\mu_{n-2}}{1\over \slash \!\!\!
q_{n-1}-z_2\slash \!\!\!\eta+m_{n-1}}\nonumber\\
%\cr 
&+& \cdots  \cdots \cdots\nonumber\\
&+&{1\over \slash \!\!\!q_1-z_{n-1}\slash \!\!\! \eta+m_1}
\gamma^{\mu_1}
\cdots \gamma^{\mu_{n-3}}{1\over \slash \!\!\!q_{n-2}-z_{n-1}
\slash \!\!\!\eta+m_{n-2}} \gamma^{\mu_{n-2}}
{m_{n-1}-\slash \!\!\!q_{n-1}+z_{n-1}\slash \!\!\! \eta\over
(q_{n-1}^2+m_{n-1}^2)}.
\label{fmomid}
\eea
A Dirac matrix inserted between two fermion propagators
corresponds to a cubic interaction vertex with a gluon field. 
For the case when this gluon field is an external one, one must insert 
external line factors and thus contract the space-time index of the
 Dirac matrix with that of the corresponding polarization vector.
If the gluon field corresponds to an internal line, then we must contract
the space-time index of the Dirac matrix with that of the gluon propagator.
At this stage it is important to stress that we are in the space-cone gauge,
such that $\eta\cdot A=0$, which means that $\slash \!\!\!\eta \slash \!\!\!
A=-\slash \!\!\!
A\slash \!\!\!\eta$. Thus in (\ref{momid}) we consider only insertions
of Dirac matrices which effectively anticommute with the null vector $\eta$.

The proof of (\ref{fmomid}) relies partly on one 
identity which we have already 
established, namely (\ref{momid}). First rewrite the fermionic propagators such that all denominators will correspond to scalar propagators.
Next cancel out in the numerator all terms with at least one  
insertion of $\slash \!\!\!\eta$. This can be done, since $\slash \!\!\!\eta 
\slash \!\!\!\eta=0$, corroborated with the previous 
observation that $\slash\!\!\!\eta$ and $\gamma^\mu$ anticommute.
Finally, we need to employ one other algebraic identity, namely
\be
\int dz \frac{z^m}{(z-z_1)(z-z_2)\dots(z-z_{n-1})}=0, \; {\rm for}\; m<n-2\,,
\ee
%\frac{1}{(z_1-z_2)(z_1-z_3)\dots (z_1-z_{n-1})}
%+\frac{1}{(z_2-z_1)(z_2-z_3)\dots(z_2-z_{n-1})}+\dots=0\,.
%\ee
where the integral is evaluated over a contour which encircles all the poles.
As mentioned many times before, the structure of the vertices is unchanged
by the shift with $z_i\eta$ in the momenta of the reference gluons.
We complete the proof of the recursion
relations by  observing that each term in (\ref{fmomid}) corresponds
to a factorization into lower on-shell amplitudes, such that
each propagator belonging on the path that connects the two reference
external gluons is cut and put on-shell, accompanied by the
corresponding shift of the external gluons. The left and right
amplitudes are multiplied by the propagator of the line which was cut. 
The reason why in the recursion relations the ``cut'' fermionic lines have
an extra factor $m_i-\slash \!\!\! q_i +z_i\slash \!\!\!\eta$ has to 
do with the fact that this corresponds exactly to the appropriate 
insertion of the external line factors. 

Thus we have provided a field theoretical proof of the recursion
relations for massive scalars and fermions found by \cite{Badger:2005zh,
Badger:2005jv}.

\section{Conclusions}
\indent 

To summarize our results, we have shown that the BCFW recursion relations 
can be proven starting from the standard gauge theory Feynman diagrams.
Perhaps the only less familiar ingredient is the use of a certain 
convenient gauge, space-cone. We have shown that each tree level
gluon Feynman diagram ``factorizes'', i.e. it can be written as a sum of
product of lower on-shell amplitudes that arise from successive cuts of 
all internal lines which connect two reference gluons. To be able 
to still satisfy the momentum conservation laws, the momenta of the reference
gluons must be shifted in a particular way. Moreover, the amplitudes
arising from cuts must be multiplied by the propagator of the line which was 
cut. The proof of factorization is based, among other things, on an algebraic
identity involving the momentum space propagators, which we recognized
as the complexified Fourier transform of the largest time equation.

Let us recall that the largest time equations are exact identities in
quantum field theories, independent of loop levels.  They yielded 
results for spectral representation of two-point functions, side-wise 
dispersion relations for vertex functions, etc.  We may infer from 
these examples that exploiting them in the area of complexified 
unitarity to complement space-cone gauge freedom should likely 
offer new opportunities at the loop level for organizing QCD 
computations, among other things. In particular, it should facilitate 
a field theoretical perspective of the one-loop recursion relations
\cite{Bern:2005hs}.

We have also addressed generalizations of the tree level BCFW recursion 
relations involving massive charged scalars and fermions, proving them in a 
purely field theoretical setup. 
\section*{Note added}

While this manuscript was being written, we became aware of 
\cite{Draggiotis:2005wq} which has partial overlap with our results.

\section*{Acknowledgments}

This work is supported in part by DOE under grant
DE-FG02-95ER40899.

%%%%%%%%%%%%%%%%%%%%%%%%%%%%%%%%%%%%%%%%

\end{document}